\documentclass{article}

\usepackage{arxiv}
\usepackage{graphicx}

\usepackage[utf8]{inputenc} % allow utf-8 input
\usepackage[T1]{fontenc}    % use 8-bit T1 fonts
\usepackage{hyperref}       % hyperlinks
\usepackage{url}            % simple URL typesetting
\usepackage{booktabs}       % professional-quality tables
\usepackage{amsfonts}       % blackboard math symbols
\usepackage{nicefrac}       % compact symbols for 1/2, etc.
\usepackage{microtype}      % microtypography
\usepackage{lipsum}
\usepackage{amsmath}

\usepackage{framed,multirow}
\usepackage{xcolor}
\definecolor{newcolor}{rgb}{.8,.349,.1}

\usepackage[ruled,vlined]{algorithm2e}
\usepackage{caption, subcaption}
\title{Automatic multiscale approach for water networks partitioning into dynamic district metered areas}

\author{
  Carlo~Giudicianni\\
  Dipartimento di Ingegneria\\
  Universit\`a degli Studi della Campania 'L. Vanvitelli'\\
  via Roma 29, Aversa 81031, Italy\\
  \texttt{carlo.giudicianni@unicampania.it} \\
  \And
  Manuel~Herrera \\
  Institute for Manufacturing -- Dept. of Engineering\\
  University of Cambridge\\
  17 Charles Babbage Rd., CB3 0FS Cambridge, UK\\
  \texttt{amh226@cam.ac.uk} \\
  \And  
  Armando~di~Nardo\\
  Dipartimento di Ingegneria\\
  Universit\`a degli Studi della Campania 'L. Vanvitelli'\\
  via Roma 29, Aversa 81031, Italy\\
  \texttt{armando.dinardo@unicampania.it} \\
  \And
  Kemi~Adeyeye\\
  Dept. of Architecture and Civil Engineering\\
  University of Bath\\
  Claverton Down, BA2 7AY Bath, United Kingdom\\
  \texttt{k.adeyeye@bath.ac.uk} \\
  \And
}

\begin{document}
\maketitle

\begin{abstract}
This paper presents a novel methodology to automatically split a water distribution system (WDS) into self-adapting district metered areas (DMAs) of different size. Complex networks theory is used to propose a novel multiscale network layout made by landmark and key nodes for the water supply plus the hyper-links representing the connection between them. The proposed multiscale layout partitioning was tested on a real medium-size water distribution network. This is shown to naturally support further DMA aggregation / disaggregation operations with the direct benefit of providing a better dynamic system control and superior efficient water management than static DMA configurations, particularly in the case of abnormal functioning conditions. The proposed tool gives the possibility to automatically define a dynamic partitioning of WDSs according to spatial and temporal water demand variability, ensuring an efficient, sustainable and low-cost management of the system whilst simultaneously preserving the hydraulic performance of the WDS.
\end{abstract}

% keywords can be removed

\section{Introduction}
Water is an important resource for population well-being and basic for a majority of the economic activities in a city. Cities can be understood as data-interconnected entities continuously growing both on complexity and population size. They are the operational centre for economic matters and well-being standards worldwide.  Water utilities face new challenges for the optimal operation and management of water distribution systems (WDSs). The proper and efficient management of a city's water infrastructure is a complex problem. In addition to supplying water to the users and satisfying the minimum service level, the main goals for water utilities nowadays range from managing abnormal conditions (such as burst pipe scenarios, peak demand variability) to dealing, at near real-time, with accidental or intentional contamination \cite{grayman2009effects,xin2017}, cyber-attacks \cite{housh2018model}, optimal sensor station placement \cite{ciaponi2018optimal}, and leakage detection \cite{wu2016burst}. Within this context, it is common practice for water networks to be split into district metered areas (DMAs) \cite{wir1999}. Water Network Partitioning (WNP) has become one of the most attractive and studied strategy for the improvement of WDS management. Over the years, working with DMAs has helped water utilities to simplify water balance computation \cite{ferrari2013graph}, carry out leakage control \cite{di2013case,taillefond2002fuzzy}, \cite{feng2006algorithm} pressure management and hydraulic performance \cite{wir1999}, monitor water quality \cite{kirstein2014simplification,chianese2017}, and speed up repairing interventions \cite{campbell2016novel}. In this way, water utilities can easily and efficiently plan management programs and compare the overall WDS performance between DMAs, by reducing the complexity of the network layout into smaller monitored areas.

Traditionally, WNP is a static solution for urban water distribution operation and management. Although working with DMAs has a number of fore-described advantages, it comes with important associated inconveniences. These are mainly related to the energy efficiency for supply and water quality. Therefore, a dynamic approach to WNP represents a valid and efficient solution for dealing with these drawbacks. This paper proposes a general framework for the dynamic partitioning of WDSs to take advantage of managing and controlling WDSs by small discrete areas at certain hours of the day (or during certain circumstances such as water disruption scenarios). The dynamic DMA expands to a topology made by a division based on larger areas compared to the initial DMAs to also take advantage of a more efficient supply at certain hours of the day (peak-demand hours) and also for maintaining regular water supply during day time. The methodology is based on the automatic definition of an optimal WNP into dynamic districts. In this way, the hydraulic performance and the resilience to failure of the water system are preserved during the day (the multiple supply paths, that are previously closed, are now open between DMAs and available immediately). The network simply reverts back to the original DMA structure at night in order to simplify the leakage detection and the management tasks. Thus, the disadvantages of a closed topology are eliminated without losing the possibility to exploit the strengths of the water network partitioning, thereby preserving its original purpose.

The current proposal pioneers an automated and practical approach to effectively achieving the dynamic water network partitioning (DWNP) related to an adaptive DMA configuration. This is achieved by a novel multiscale abstraction of the original water network layout. The process continues by approaching a clustering algorithm that takes into account the current DMA division. Consequently, the initial smaller DMAs are dynamically: a) aggregated (grouped) into bigger areas by a clustering methodology that preserves the connectivity of the original water network, in order to preserve the network resilience, improve the pressure management, ensure the water quality; and b) periodically desegregated (i.e. each night, once per week, etc.) according to the specific objective of the water utilities (i.e. leakage monitoring at night).

The main objective of the proposed research is to develop a generic framework for the dynamic top-down / bottom-up partitioning of WDSs for a smart and efficient management in response to abnormal functioning conditions. A DWNP tool is particularly tailored for pressure control, leakage detection and contaminant spreading stop, at different stages (night / morning hour, winter / summer time, normal / abnormal conditions). In particular, in this paper, the framework has been applied to address non anticipated increase in the peak demand. While in the last decade, several methodologies for WDS partitioning has been widely studied \cite{perelman2015automated}, dynamic strategies on which this work is based, still represents a novel fieldwork in urban hydraulics. Few works can be considered antecedents of this paper; notably the work of \cite{wright2014adaptive} which proved great advantageous when such dynamic approaches are applied to real networks for a DMA based management \cite{wright2014dynamic}, and \cite{di2016dynamic} which pointed out the importance of realizing smart WDS by using remote-controlled valves. However, these works only investigates the benefits of dynamic partitioning but falls short of providing a methodological approach to actually deliver it.

\section{Permanent water network partitioning}

The management of a WDS through its division into DMA is based on defining areas partially isolated from the rest of the network by the insertion of gate-valves and flow-meters along some pipes. WNP is usually accomplished in two phases \cite{perelman2015automated,di2017economic}:

\begin{enumerate}
    \item $clustering$, aimed to define the shape and the dimension of the network subsets, to balance the number of nodes of each cluster and to minimise the number of edge-cuts (boundary pipes) $N_{ec}$, using graph algorithms \cite{deuerlein2008decomposition,perelman2011topological,di2010design}, multilevel partitioning \cite{alvisi2015new}, community structure \cite{diao2012automated}, multi-agent systems \cite{herrera2010water}, spectral approaches \cite{herrera2012multi,di2018applications}, modularity-based procedure \cite{ciaponi2016modularity}
    
    \item $dividing$, aimed to the physical division of the network, by selecting pipes along which flow meters or gate valves have to be inserted, minimising the economic investment and the hydraulic performance deterioration, based on iterative approach \cite{ferrari2013graph} or optimisation algorithms \cite{di2017economic}, with the objective of defining the optimal layout that minimises the cost and the hydraulics deterioration \cite{izquierdo2009division,gomes2013district}.
\end{enumerate}

Defining a WNP is a real challenge for operators as it depends on the design choices at both clustering and partitioning phases. A permanent partitioning of WDSs (by extensively closing boundary valves partitioning the network) increases the economic investment and could lead to inefficient supply. This is since the more partitioned the network the more dissipated the supply energy and so lower water pressures at the end user \cite{wright2014adaptive}. This might also deteriorate the hydraulic performance and reliability of the system due to the closure of a number of boundary pipes. This is because the partitioning of the water network reduces the natural redundancy in connectivity of the water systems \cite{herrera2016graph}.

The optimal WNP design and the related hydraulic simulations are often carried out by referring to ordinary operation condition (e.g. peak water demand), based on a static representation of the system. However, WDS could face extraordinary conditions (i.e.unplanned water demand peaks, insufficient pumping or storage capacity, fire, pipe breaks) that strongly compromise the performance of the system. The importance of taking into account the spatio-temporal variability of water request across the WDS was highlighted in \cite{di2018performance} where the Authors tested a partitioned WDS under random spatio-temporal variability of peak demand. They showed that the water request variability strongly affects the hydraulic performance of the system and it should be taken into account during the design of network partitioning. Another issue when working with permanent DMAs is related to water ageing, particularly at lower consumption network areas. This would require timed manual consumption interventions (i.e. opening of closed boundary valves where stagnant water have accumulated, leaving valves open or closed inappropriately) without which water safety and suitability cannot be guaranteed. 

Despite the advantages of the control and monitoring of WDS by using permanent DMAs on leak detection, there also are some disadvantages that should be taken into account for an optimal WNP. Table~\ref{tab:pros_and_cons} summarises the pros and cons of a permanent WNP.

\begin{table}[!ht]\scriptsize
\caption{Advantages and disadvantages of a permanent Water Network Partitioning} 
\label{tab:pros_and_cons} 
\centering 
\begin{tabular}{l l}
\hline
\textbf{Advantages}  & \textbf{Disadvantages} \\
\hline
  Improved leakage detection and identification & Reduced resilience to failures \\
  Simplified sub-area management & Reduced operational flexibility \\
  Improved protection against contaminant & Water issues in peripheral areas \\
  Improved sub-area pressure control & Investment and maintenance costs \\
\hline  
\end{tabular}
\end{table}

The target of a partitioning layout design is to balance the negative and positive aspects described in Table \ref{tab:pros_and_cons}, ensuring the fulfilment of the minimum required nodal pressure, so as to satisfy users demand and preserve network resilience.

%%%%%%%%%%%%%%%%%%%%%%%%%%%%%%%%%%%%%%%%%%
\section{Dynamic water network partitioning}

This section introduces the process for dynamic water network partitioning. This is approached through the novel concept of multiscale (MS) water network layout. The first advantage of this network abstraction is on network visualisation. This is specifically useful for large water networks. In addition, the MS water network is shown to be useful for automating the creation of dynamic DMA by a semi-supervised MS clustering.

\subsection{Multiscale network layout}
Creating a MS network layout starts with a network already divided into clusters or communities. There are the following key components encompassing the MS approach:

\begin{itemize}
    \item $single$ $assets$: links and nodes representing generic elements of the original network. In the case of WDSs, these are: demand nodes, valves, junctions, pipes, tanks, reservoirs, and pump stations;
    \item $boundary$ $node$: (landmark) nodes acting as inlet / outlet of a network previously divided into discrete metered areas;
    \item $hyper$-$links$:
     \begin{itemize}
        \item $boundary$ $links$: links connecting boundary nodes belonging to different clusters or DMAs;
        \item $internal$ $links$: links connecting boundary nodes belonging to the same clusters or DMAs;
    \end{itemize}
\end{itemize}

Following this network decomposition, it is possible to make a MS network directly related to the original layout but only compressing the items that are key for its connectivity. In addition to providing a novel visualisation of the original network, the approach eases further connectivity analysis at cluster (or community) level.

\begin{figure}[!ht]
    \centering
    \begin{subfigure}[t]{0.48\textwidth}
        \centering
        \includegraphics[width=\textwidth]{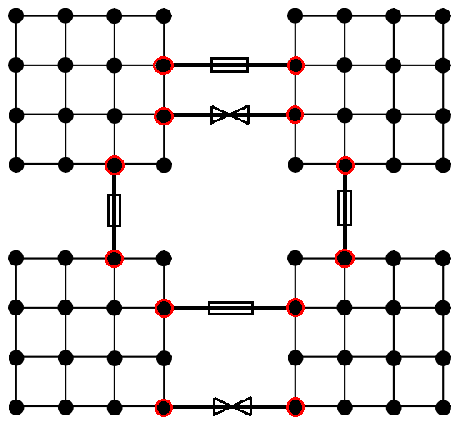}
        \caption{Toy example of network showing key elements of a WDSs: demand and boundary nodes, gate valves, flow meters, pipes.}
        \label{fig:MS-net1}
    \end{subfigure}%
    \hspace{15mm}
    \begin{subfigure}[t]{0.35\textwidth}
        \centering
        \includegraphics[width=\textwidth]{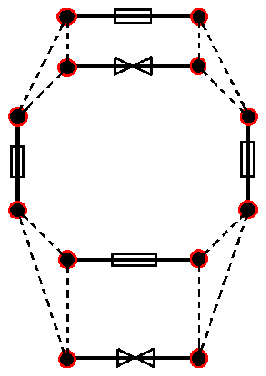}
        \caption{MS network showing key elements of the new network}
        \label{fig:MS-net2}
    \end{subfigure}
    \caption{Graphical explanation for the creation of the MS network layout}
    \label{fig:MS-net} 
\end{figure}

Figure \ref{fig:MS-net} shows how the MS network decomposition works. Figure \ref{fig:MS-net1} is a hypothetical example of a water network split into 4 straightforward DMAs. The MS network is in Figure \ref{fig:MS-net2} where the main novelties of this decomposition can be observed, as it uses the boundary nodes (black and red circles) as unique nodes remaining from the original configuration of the network, boundary links (bold black lines) and internal links (dashed black lines). 

\subsection{Dynamic district metered areas}
Many water utilities already work with a WDS divided into DMAs. Starting from this stage, larger DMAs are created by the grouping of current DMAs. An initial idea is to simply use the original WDS labelled depending on the DMAs. Then it is possible to reorganise these labels in a lower number of DMAs. This preliminary approach is straightforwardly improved by using the corresponding MS network associated to the original WDS. The main items of the MS network are the landmark nodes. In our case, these landmark nodes are boundary nodes connecting one DMA to another. This is vital information for grouping DMAs but also to maintain control within the water supply at any stage of the process as it tracks all the DMAs inlets / outlets. The boundary nodes are connected between themselves by valves and pipes, represented as boundary links. The internal hyper-links represent the connectivity within a DMA. These are important to preserve the MS network connectivity. In addition, they provide information of the internal connectivity after any DMA aggregation process.

The hyper-links between DMAs are weighted by the strength of their connection. The weight can be proposed as the shortest path but other approaches are possible. For instance, by considering the diameter and length of pipes. Boundary nodes and hyper-links create a novel MS network with all the information necessary for grouping DMAs into bigger areas. 

\subsection{Semi-supervised clustering algorithm}

The aggregation process is done by applying semi-supervised clustering \cite{kulis2009semi} running over the MS network and taking into account features such as the boundary nodes membership and internal DMA connectivity as constraints. In this way, the topology of the network and the key hydraulic nodes (inlet / outlet) of the different DMAs are taken into account. In particular, this structural knowledge comes in the form of pairwise must-link (boundary links) and cannot-link (internal links) constraints.  

It is worth highlighting that a semi-supervised approach is particularly suitable for working with real-world systems since some background knowledge about the structure and the behaviour are available \cite{herrera2010}. In this case, the algorithm have to assure that: 

\begin{itemize}
  \item the new aggregate DMAs include the former districts without splitting them
  \item the previous cluster layout and the devices already installed are exploited 
  \item the set of new boundary links is included in the set of boundary links of the original partitioning
\end{itemize}

\noindent in order to better manage the aggregation / disaggregation phases, minimise / nullify new investment costs and to reduce the computational burden of the whole procedure, providing a simple automatic dynamic schedule of the WDS. These hydraulic / structural features have to be translated as constraint implementation in the multiscale clustering algorithm. 

This paper makes a MS network according to the connectivity of the original water distribution network and shows that the structural knowledge is implicitly taken into account and constraints are respected. In fact, the MS approach produces a shift in the structure of the MS network which can be regarded as low interconnected small-world clusters (whose links are the $internal$ $links$). In fact, after the size reduction provided by the application of the MS algorithm, each cluster of the MS network becomes a fully connected layout connected to each other by fewer links ($boundary$ $links$). Thanks to this topological properties of the MS network (see Figure \ref{fig:MS-net2}), it is ensured that the clustering algorithm, applied for the definition of new bigger clusters, provides a solution in which the novel set of boundary links is a sub-set of the boundary links of the original cluster layout. On top of this, a network community detection algorithm \cite{fortunato2016community} splits a network in such way that each cluster is formed by elements having a high density connection between each other and a lower probability to be connected to items belonging to other clusters. According to this criterion, the new cluster layout will certainly cross the former boundary links and will not split the original DMAs. A semi-supervised approach automatically takes into account these particular issues by the structure of the problem itself without the necessity to build new features vector or matrix, as it is usual for this process. Compared to other methodologies, which only use graphical or vector information, semi–supervised clustering use both, and in a more efficient and robust way. The flexibility to take into account different inputs into the study, in an automatic way, exploiting the topological features of the MS network is a novel improvement of the shown methodology. 

Finally, this paper explores the widely used community detection algorithm introduced by \cite{Girvan2002}, based on the search of the edges that are most ``between'' communities. In particular, the Girvan-Newman algorithm is based on a generalisation of the betweenness centrality \cite{Freeman1977}, with a complexity of $O(n^3)$. The betweenness centrality $b(i)$ of a node $i$ is defined as the number of shortest paths between pairs of other vertices that run through $i$ (it is a measure of the influence of a node over the flow of information / substance between other nodes). Girvan-Newman extends the concept to edges, defining the edge betweenness $bc(l)$ of an edge $l$ as the number of shortest paths between pairs of nodes that run along it, in order to find which edges in a network are most between other pairs of nodes. In this way, if a network contains communities that are only loosely connected by a few inter-communities edges, then all the shortest paths between different communities have to pass along one of these few edges. According to this criterion, the edges connecting communities will have high edge betweenness $bc(l)$, and by removing these edges, the communities are separated one from another.

\subsection{Dividing phase}
The clustering phase provides the cluster layout (size and shape of each cluster) and the set of boundary link $N_{ec}$ between clusters (i.e. the set of pipes along which gate valves or the  flow meters, installed or to install). If $N_{fm}$ is the number of flow meters, $N_{gv} = N_{ec} - N_{fm}$ corresponds to the number of gate valves (e.g. closed pipes). Generally, the main goal is to keep as lower as possible the number of flow meters, $N_{fm}$, to simplify the water budget computation \cite{di2017economic}. The number of all the possible dividing configurations $N_{dc}$ (which pipes are closed and which of them are open) is expressed by the binomial coefficient at the Equation \eqref{eq:binom}.

\begin{equation} \label{eq:binom}
    N_{dc} = \binom{N_{ec}}{N_{fm}} = \frac{N_{ec}!}{N_{fm}!N_{gv}!}
\end{equation}

A classical computational problem is that the number $N_{dc}$ is so huge that it is often computationally unmanageable to investigate all the space of solutions. This also occurs for small-size water distribution system. This is the reason why a heuristic optimisation approach has been adopted to find the optimal position of physical assets such as flow meters and gate valves on the boundary links. In this case, a Genetic Algorithm (GA) \cite{goldberg1988genetic} was developed and specifically tailored for this problem. The Objective Function (OF) of Equation \eqref{eq:GA_obj} was maximised.

% The GA (developed in Python3.7)

\begin{equation} \label{eq:GA_obj}
   I_r = \frac{\sum\limits_{i=1}^{n_n} Q_i(h_i-h^*)}{\sum\limits_{r=1}^{n_r} Q_rH_r - \sum\limits_{i=1}^{n_n} Q_ih^*}
\end{equation}

Equation \eqref{eq:GA_obj} corresponds to the resilience index $I_r$ of \cite{todini2000looped}; where $n_n$ is the number of demand nodes, $n_r$ is the number of reservoirs, $Q_i$ and $h_i$ are respectively the water demand and the pressure head of the i-th node, $Q_r$ and $H_r$ are the water discharge and the total head of the generic $r$-th source point, and $h*$ the design pressure head of the network (i.e. the minimum required pressure to guarantee the minimum service level to the users).

The GA was carried out with 100 generations of a population consisting of 50 individuals. Each individual of the population is a sequence of a number of binary chromosomes equal to the number of boundary links $N_{ec}$ and corresponding to them. Each chromosome assumes value 0 if a gate valve is inserted in the j-th pipe, value 1 otherwise if a flow meter is installed. The crossover percentage is settled $P_{cross}$ = 0.8, and the mutation rate $P_{mut}$ = 0.02. 
Further, the OF is constrained by the expression of Equation \eqref{eq:hmin}, which imposes a minimum service level for all the users:

\begin{equation} \label{eq:hmin}
     h_{min} \ge h^* 
\end{equation}

\noindent where $h_{min}$ is the minimum nodal head pressure.

Among the set of optimal solutions the one also minimising the number of flow-meters has been chosen. The criterion is such that the smaller the number of flow-meters, the simpler the water budget computation and the WDS management.

\subsection{Decision Support System}
\begin{figure}[!ht]
    \centering
    \includegraphics[width=\textwidth]{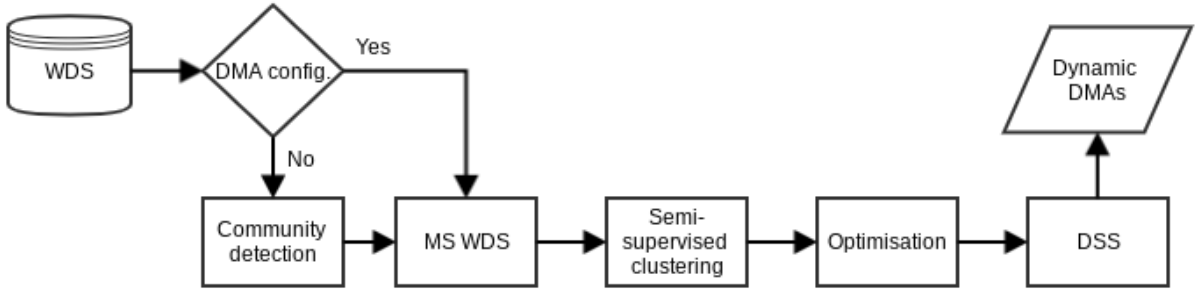}
    \caption{Flowchart of the process for the creation of Dynamic DMAs}
    \label{fig:Flowchart_DDMA} 
\end{figure}

The proposed framework provides the possibility to also take into account an economic criterion for selecting a final optimal solution between the set of configurations that satisfied the hydraulic criteria (resilience index and minimum head pressure). Dynamic DMAs aid any Decision Support System (DSS) adopted by water utilities. This is achieved by a further selection of the partitioning configurations that completely exploit the devices already installed over the water distribution system, with no further need to buy new flow meters / gate valves. In these case, the dynamic aggregation / disaggregation of the districts requires only the opening and closing of valves, and strongly simplifies the management of scenarios of unexpected emergencies. Thus, the procedure is able to simultaneously take into account the hydraulic and economic criteria and to speed up the computational operation for finding out the optimal solution. The overall process on the creation of Dynamic DMAs is summarised in Figure \ref{fig:Flowchart_DDMA}.

%%%%%%%%%%%%%%%%%%%%%%%%%%%%%%%%%%%%%%%%%%
\section{Experimental study}

The creation of the dynamic DMAs is tested in a real, medium-sized WDS. An abnormal water request by users that worsens the hydraulic performance of the WDS is simulated, nullifying the advantages of the previous partitioning. This shows the benefits of working with dynamic DMAs for a water utility as according to stressing operational conditions, a new DMA configuration should be required. The new dynamic WDS partitioning should ensure the optimal operation and management of the system whilst restoring the minimum service level for the users.

\subsection{Case-study of Parete WDS}
The proposed methodology is tested on a real medium-size WDS serving the city of Parete (Italy). Parete WDS currently serves a population of more than 11,000 inhabitants. As Figure~\ref{fig:Parete11} shows, this network encompasses 182 junctions and 2 reservoirs, with fixed head of 110 m a.s.l, ($n = 184$ nodes), and 282 pipes ($m = 282$ links). 
The hydraulic performances of the un-partitioned WDS of Parete, reported in Table~\ref{tab:topo_hydra}, is good in terms of the maximum $h_{max}$, mean $h_{mean}$ and minimum $h_{min}$ pressure head (higher than the design pressure head $h^*$ = 19 m assumed for all the demanding nodes). In particular, $h^*$ = 19 m comes from the sum of the maximum building height in the town (which is 9 m in Parete), plus 10 m, as prescribed by the Italian guidelines. A previous water network partitioning was carried out in order to simplify the management of the system and the leakage detection. In particular, the WDS of Parete is partitioned into $C=9$ DMAs, and the district layout is shown in Figure~\ref{fig:Parete11}. The partitioning is made according to the day of maximum consumption in the year when the total nodal demand ranges from 7.6 L/s at night time to 110.2 L/s in the morning and midday peaks, with an average value of 36.3 L/s. 

The main topological and hydraulic characteristics together with the simulation results of the WDS are reported in Table~\ref{tab:topo_hydra}; where $I_b$ is the cluster balanced index (it corresponds to the standard deviation and quantifies how well the clusters are balanced in terms of the number of nodes). It is clear that, the cluster layout corresponding to $9 DMAs$ $(p=3.00)$ is well balanced, having $I_b$ = 4.45, the number of flow meters is kept low $N_{fm}$ = 13 (considering that the number of boundary pipes is $N_{fm}$ = 20) and the hydraulic criteria are also all satisfied (respectively $h_{max}$ = 50.27 m, $h_{mean}$ = 28.23 m, $h_{min}$ = 19.05 m and $I_r$ = 0.369), with a good value of resilience index (a reduction of around 23\%), if compared with the un-partitioned layout ($Un-partitioned$).

The experimental study continues by simulating a quite strong urbanization all over the city, leading to an increase in water demand by the users, shifting the water demand up to 15\% during the midday peak and for the whole WDS. Without loss of generality, it is assumed that the water sources are able to address this increasing request. This scenario leads to a strong decrease of the performance of the WDS and, particularly, its resilience. Under these conditions, the 9 DMAs partitioning does not satisfy the design criteria as described above (see $9 DMAs$ $(p=3.45)$). As it is evident in Table~\ref{tab:topo_hydra}, the WDS is now characterised now by a lower value of pressure head (respectively $h_{max}$ = 49.08 m, $h_{mean}$ = 22.55 m, $h_{min}$ = 10.87 m) and a resilience index $I_r$ = 0.168 (a reduction of around 65\% with respect to the $Un-partitioned$ layout). In order to address this problem (managing the difference in performance between the normal and abnormal water request conditions), a dynamic aggregation / disaggregation of DMAs seems to be a valid solution. This preserves both the advantages of the water network partitioning and delivers a suitable resilience level of the system.

\begin{table}[!ht]\scriptsize
\caption{Topological and hydraulic characteristics of the original un-partitioned WDS, of the network partitioned in C=9 DMAs during normal midday peak, and the network partitioned in C=9 DMAs during the abnormal midday peak for Parete WDS} 
\label{tab:topo_hydra} 
\centering
\begin{tabular}{l c c c c c c c c}
\hline
\textbf{Layout}  & \boldmath{$N_{ec}$} & \boldmath{$N_{fm}$} & \boldmath{$N_{gv}$} & \boldmath{$I_{b}$} & \boldmath{$h_{min}$} & \boldmath{$h_{mean}$} & \boldmath{$h_{max}$} & \boldmath{$I_{r}$} \\
& [-] & [-] & [-] & [-] & [m] & [m] & [m] & [-] \\
\hline
  Un-partitioned & -  & - &  - & - & 21.36 & 31.05 & 50.47 & 0.481 \\
  9 DMAs (p=3.00) & 33 & 13 & 20 & 4.45 & 19.05 & 28.23 & 50.27 & 0.369 \\
  9 DMAs (p=3.45) & 33 & 13 & 20 & 4.45 & 10.87 & 22.55 & 49.08 & 0.168 \\
\hline  
\end{tabular}
\end{table}

\subsection{Dynamic DMA configuration for Parete WDS}

\begin{figure}[!ht]
    \centering
    \begin{subfigure}[t]{0.40\textwidth}
        \centering
        \includegraphics[width=\textwidth]{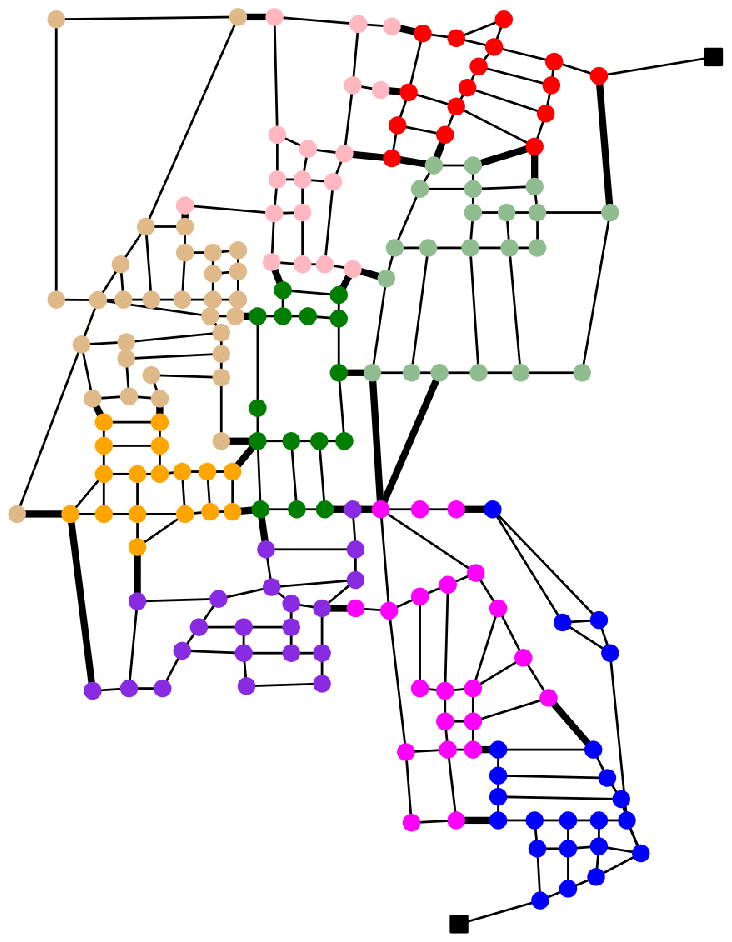}
        \caption{DMA division for Parete WDS}
        \label{fig:Parete11}
    \end{subfigure}%
    \hspace{15mm}
    \begin{subfigure}[t]{0.40\textwidth}
        \centering
        \includegraphics[width=\textwidth]{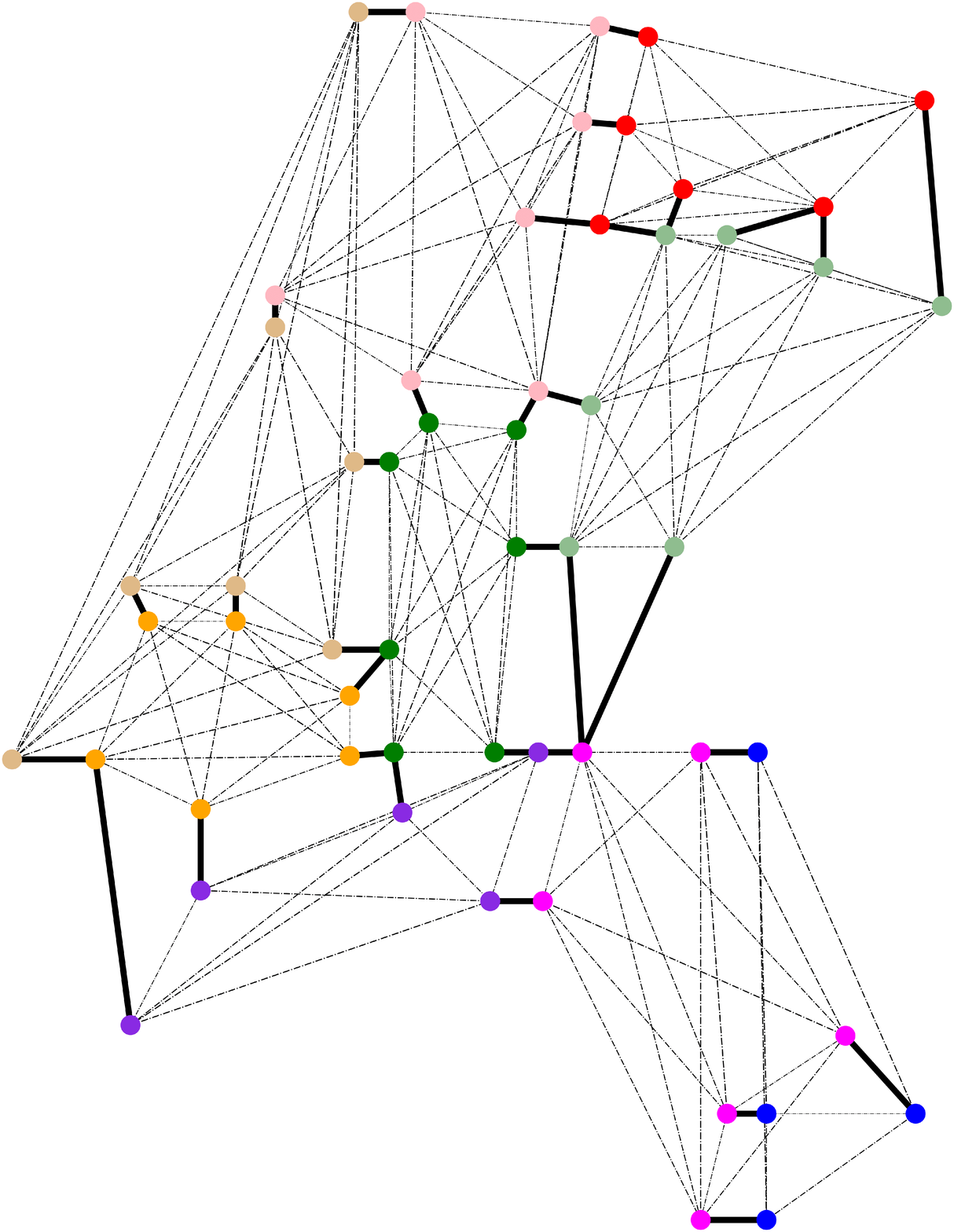}
        \caption{MS Parete WDS}
        \label{fig:Parete12}
    \end{subfigure}
    \caption{Graphical explanation for the creation of Parete MS WDS layout (9 DMAs)}
    \label{fig:MS-Parete1} 
\end{figure}

First, the proposed dynamic DMAs configuration detects the clustering layout of the WDS and then builds the corresponding MS network. Figure~\ref{fig:Parete12} shows the size reduction of the Parete WDS after its transformation into a MS network. This figure also shows the key elements, such as the boundary nodes of each cluster (highlighted by their corresponding DMA colour at Figure~\ref{fig:Parete11}), the boundary links (bold black line), and the internal links (thin dashed grey line). The set of hyper-links (boundary links and internal links) is crucial for automatically implementing the semi-supervised clustering during the aggregation phase. Boundary links represent the connectivity between different clusters, while the internal links constitute the internal connectivity of each cluster (according to the shortest path linking each pairs of boundary nodes belonging to the same cluster).

\begin{figure}[!ht]
    \centering
    \begin{subfigure}[t]{0.40\textwidth}
        \centering
        \includegraphics[width=\textwidth]{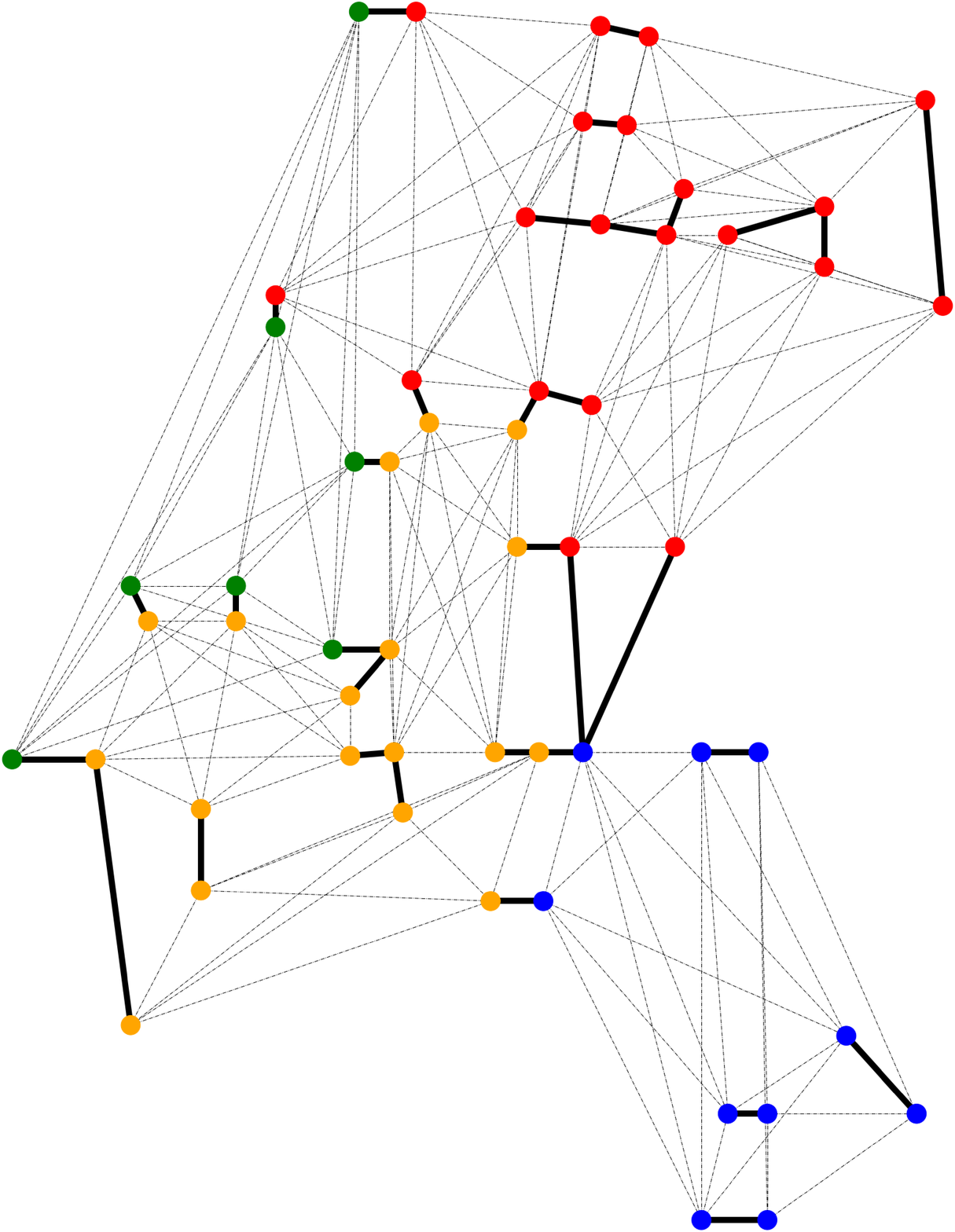}
        \caption{DMA aggregation over MS Parete WDS}
        \label{fig:Parete21}
    \end{subfigure}%
    \hspace{15mm}
    \begin{subfigure}[t]{0.40\textwidth}
        \centering
        \includegraphics[width=\textwidth]{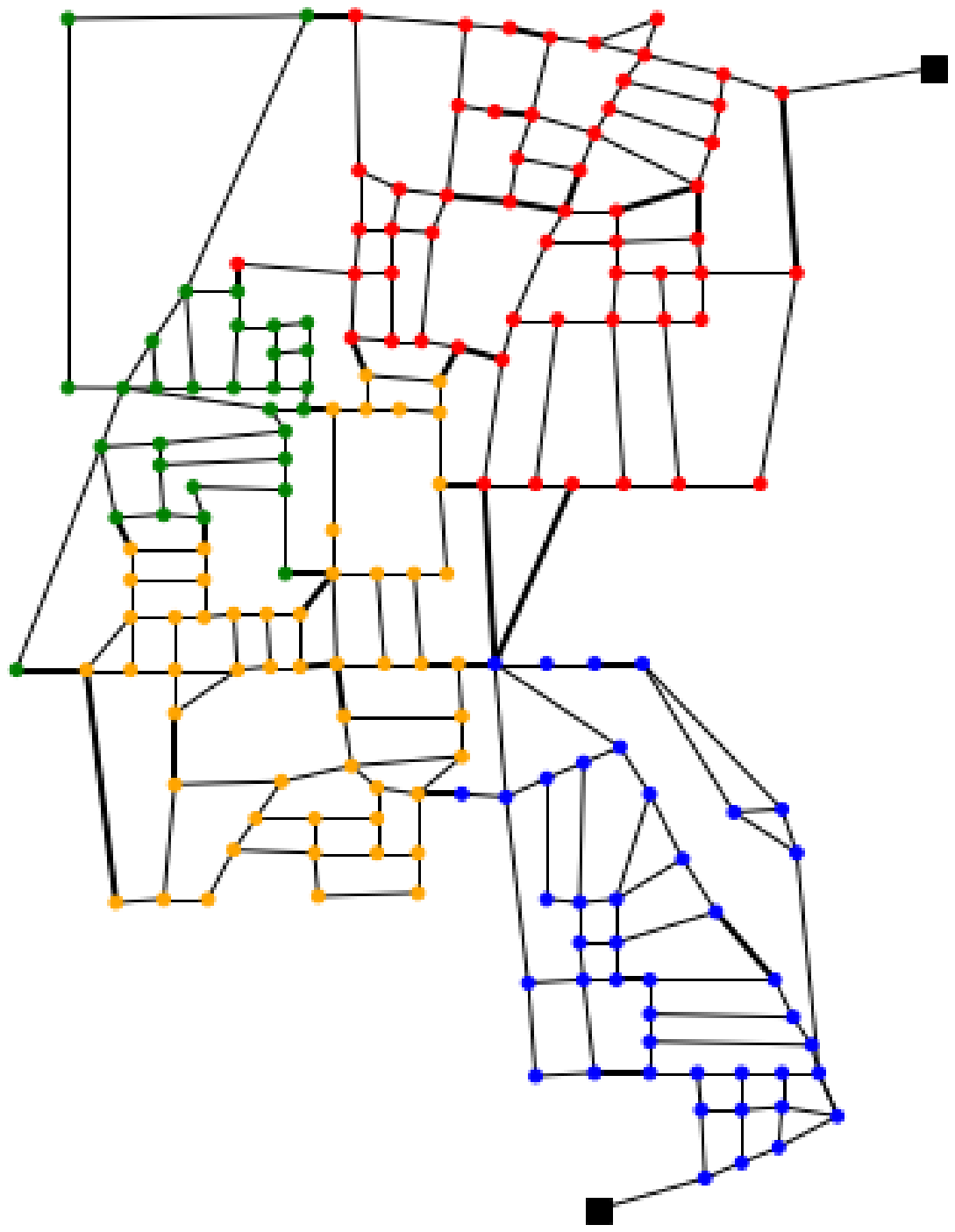}
        \caption{DMA division for Parete WDS}
        \label{fig:Parete22}
    \end{subfigure}
    \caption{Graphical explanation for the creation of Parete MS WDS layout (4 DMAs)}
    \label{fig:MS-Parete2} 
\end{figure}

A suitable number of clusters $C$ is taken in order to optimise the overall connectivity of the partitioned network. According to \cite{Giudicianni2018}, the optimal number of clusters (from a topological point of view) results in $C_{opt} \propto n^{0.28}$. Thus, without loss of generality, the new number of clusters for Parete is set to $C=4$ (see Figure~\ref{fig:Parete21}). The first step of the proposed methodology is to aggregate the previous DMAs in the MS network, satisfying all the criteria associated with the semi-supervised clustering. The Girvan-Newman algorithm is applied to the MS network to provide the new clustering layout which suitably balances the new bigger 4 DMA (in terms of number of nodes) and minimises the number of boundary links between clusters. These are two crucial aspects for the definition of the new clustering configuration, since they ensure a better management (district with same size) and reduce the computational burden in the subsequent dividing phase: a lower number of boundary links $N_{ec}$ leads to a lower number of dividing configuration $N_{dc}$, as evident in \eqref{eq:binom}. Another important aspect to highlight is that the application of the clustering phase on the MS network drastically reduces the computational time since as it was previously stated, the community structure algorithms generally has a complexity proportional to the number of nodes $n$. 

Figure~\ref{fig:Parete22} shows that the 4 new, bigger clusters perfectly include the former DMAs (avoids to split them), and so the new DMA configuration respects the internal continuity. They are well balanced (as listed in Table~\ref{tab:topo_hydra_1}, the balance index is $I_b$ = 9.72) and finally the new set of boundary links $N^*_{ec}$ = 14 constitutes a subset of the previous set $N_{ec}$ = 33 (e.g. all the new boundary links were boundary links also for the partitioning in $C=9$ DMAs). 

This represents the most important point of the dynamic aggregation / disaggregation phase since it ensures that the DMAs in each phase are kept in control, exploiting the devices already installed in the WDS. Finally, it is important to note that from the MS network is possible to return to the original layout of WDS by exploding the MS characteristics and layout. Figure~\ref{fig:Parete22} shows all the nodes in each DMA configuration visualised by different colours. 

\begin{table}[!ht]\scriptsize
\caption{Topological and hydraulic characteristics of the network partitioned in $C$=4 DMAs during the abnormal midday peak for Parete WDS} 
\label{tab:topo_hydra_1} 
\centering 
\begin{tabular}{l c c c c c c c c}
\hline
\textbf{Layout}  & \boldmath{$N^*_{ec}$} & \boldmath{$N^*_{fm}$} & \boldmath{$N^*_{gv}$} & \boldmath{$I_{b}$} & \boldmath{$h_{min}$} & \boldmath{$h_{mean}$} & \boldmath{$h_{max}$} & \boldmath{$I_{r}$} \\
& [-] & [-] & [-] & [-] & [m] & [m] & [m] & [-] \\
\hline
  4 DMAs (p=3.45) & 14 & 5 & 9 & 9.72 & 16.21 & 25.61 & 48.93 & 0.274\\
  4 DMAs (p=3.45)-GA & 14 & 7 & 7 & 9.72 & 17.56 & 25.98 & 49.24 & 0.303\\
\hline  
\end{tabular}
\end{table}

After the definition of the new clustering layout, the boundary links inside the new 4 bigger DMAs are opened ($4 DMAs$ $(p=3.45)$ layout) and the check of the hydraulic performance is made. This step ensures that the management of the aggregation / disaggregation phase is further simplified. If the minimum service level is not yet satisfied (as it happens in the analysed case, for which $h_{min}$ = 16.21 m and the resilience index is $I_{r}$ = 0.274, as reported in Table~\ref{tab:topo_hydra_1}), the dividing phase, with the GA optimisation, is carried out on the new DMAs configuration, according to the Equation \eqref{eq:GA_obj}. 

\begin{figure}[!ht]
    \centering
    \includegraphics[width=\textwidth]{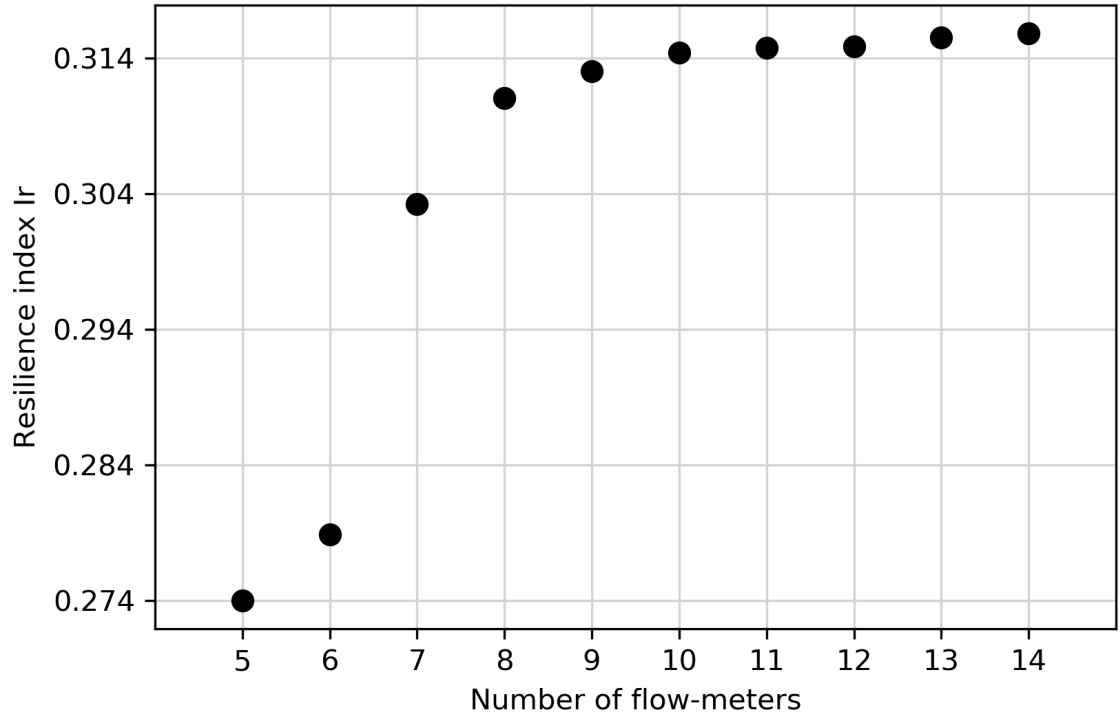}
    \caption{Decision Support System for the optimal partitioning layout of Parete WDS}
    \label{fig:DSS} 
\end{figure}

Of all the possible optimal solutions provided by the optimisation, those that exploit the already installed flow-meters on the WDS, are chosen as the best ones to further consider at the DSS stage. In this way, it can be ensured that solutions are also convenient from an economic point of view, as only the installation of some of the necessary flow-meters will be required. The DSS is reported by Figure~\ref{fig:DSS}, in which layouts providing a maximum number of 14 flow-meters are reported (the latter case corresponds to virtual partitioning, for which on all the boundary pipes are installed flow-meters). Furthermore, in order to better compare the different layouts, also the solution $4 DMAs$ $(p=3.45)$ with 5 flow-meters is reported.  

A possible choice in this context is the solution with 7 flow meters (as consequence the number of gate valves $N^*_{gv}$ = 7) as the most suitable configuration for the case study. The simulation results are shown in Table~\ref{tab:topo_hydra_1}, from which it is clear that the resilience of WDS for the $4 DMAs$ $(p=3.45)-GA$ solution is restored ($I_{r}$ = 0.303, just 17\% less than that of the original partitioned layout $9 DMAs (p=3.00)$, and 4\% less than the resilience of the virtual partitioning with $N^*_{fm}$ = 14). Regarding the minimum pressure $h_{min}$ = 17.56 m is slightly lower than $h*$=19 m (but only 9 nodes do not satisfying the pressure constraint). It is worth mentioning that the water utility can choose another solution according to its need and budget and can even automate the overall decision-making process. 

\begin{figure}[!ht]
    \centering
    \begin{subfigure}[t]{0.40\textwidth}
        \centering
        \includegraphics[width=\textwidth]{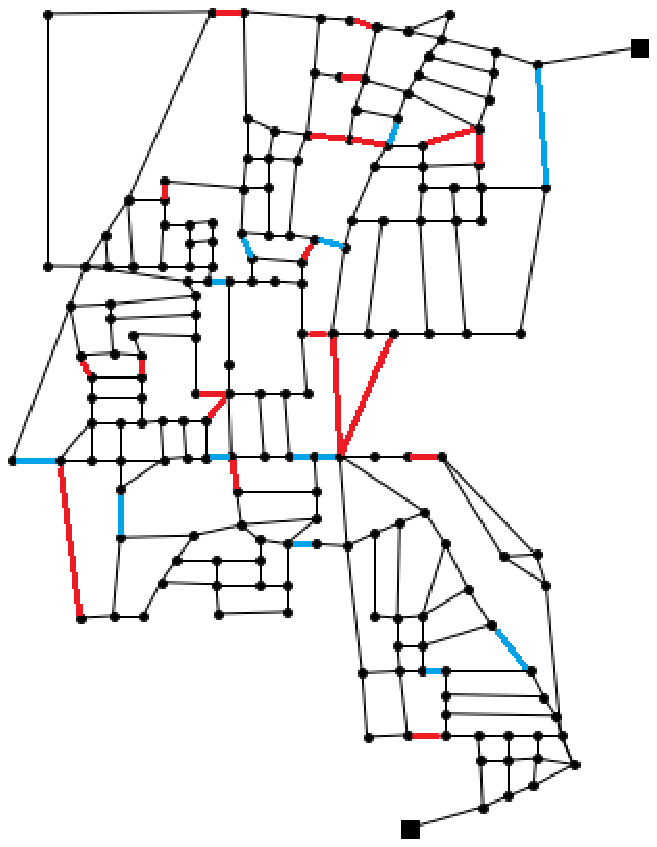}
        \caption{9 DMAs}
        \label{fig:9dmas}
    \end{subfigure}%
    \hspace{15mm}
    \begin{subfigure}[t]{0.40\textwidth}
        \centering
        \includegraphics[width=\textwidth]{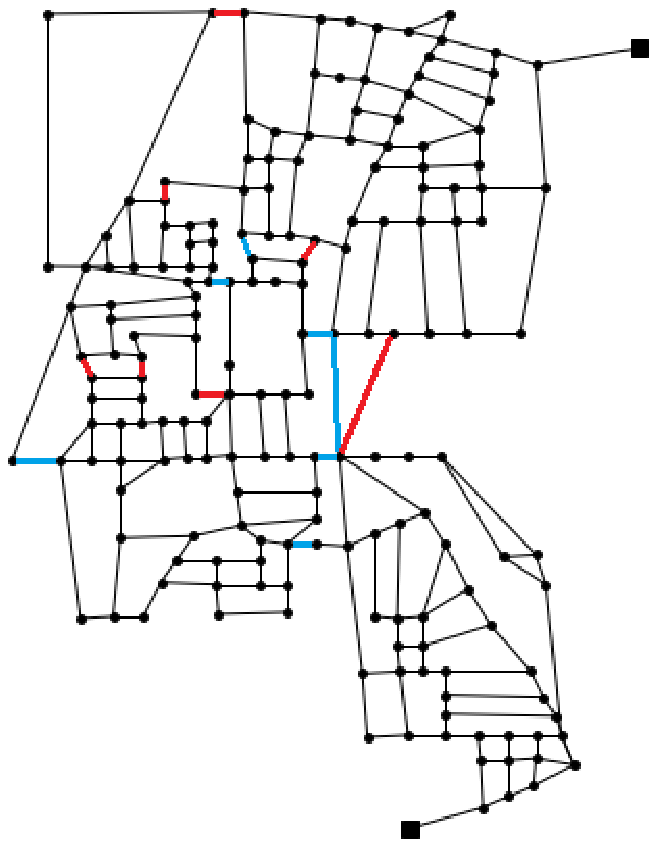}
        \caption{4 DMAs}
        \label{fig:4dmas}
    \end{subfigure}
    \caption{Optimal partitioning layout for Parete WDS; red lines for gate-valves and blue lines for flow-meters}
    \label{fig:dmas} 
\end{figure}

Figure~\ref{fig:dmas} shows the two partitioning layouts. Specifically, Figure~\ref{fig:9dmas} shows the previous partitioning in 9 DMAs, while Figure~\ref{fig:4dmas} shows the best solution composed by 4 bigger DMAs (after the optimisation phase). It is evident that all the boundary pipes of the new partitioning layout constitute a sub-set of the previous boundary pipe set. In addition, five of the seven installed flow-meters are the same of those already installed for the partitioning in 9 DMAs (blue lines).  

%%%%%%%%%%%%%%%%%%%%%%%%%%%%%%%%%%%%%%%%%%
\section{Conclusions}
This paper proposed a novel tool for an optimal dynamic partitioning of water distribution systems (WDSs) under abnormal not-contemplated conditions (such as peak demand increase). By combining the graph theory features of WDSs and advances in modelling and optimisation, a new strategy of sustainable management was proposed. This can be used to make WDSs dynamic, resilient, smarter and more adaptive in response to functionality changes and incidents. 

The dynamic network partitioning of a WDS into district metered areas (DMAs) is based on the definition of a size-reduced multiscale (MS) graph network of the WDS, which preserves only some key items of the original layout. The new graph reconfiguration simplify the top-down / bottom-up clustering, and always preserve the set of boundary pipes at each level. This ensures that for any new DMA layout the boundary set always constitutes a sub-set of the original one. In this way, all the solutions provided by the new framework, not only balance the cluster size and minimise the number of boundary links but also guarantee a low cost partitioning layout, since there devices are already installed. 

The dynamic DMAs approach is tested on a real medium-size WDS serving the town of Parete (Italy). The simulation results show its efficiency in providing an optimal sustainable solution which simultaneously ensures the restoration of the hydraulic performance, after the increase of water request, and also reduces the investments cost of a new partitioning reconfiguration. The methodology is completely automated. Furthermore, the MS graph reduces the computational burden of the problem of water network partitioning, making the proposal particularly suitable for large WDSs. The problem faced in this paper on the increment of peak-demand also stresses the importance of taking into account the spatio-temporal variability of the water request during the definition of the optimal water network partitioning. In fact, as it is shown in this paper, just an increase of 15\% in the water demand strongly deteriorates the performance of the system. 

Water demand at scenarios do not homogeneously vary over the whole network and many issues are foreseen to happen (and should be tackled) at sector level. This is the case of city areas prone to suffer flooding events, disparate differences on pipe age between network areas, or just customer type per DMA with domestic water use (also varying on family size, occupancy levels, house envelope, e.g.), public water use (official buildings, hospitals, schools), or industrial water use. This last is the most invariant DMA with respect to the water demand and exogenous conditions. Future work will address the automatic reconfiguration of the dynamic DMAs at partial network level. That is, DMAs of a certain sub-graph of the WDS are involved in the automatic reconfiguration while the rest of DMAs remain unchanged. This will help to take preventive and mitigation actions facing abnormal conditions such as the spread of contaminants, the failure of multiple pipes, extreme weather events, or fire fighting. The future aim will be to extend the multiscale dynamic DMA framework presented herein to open a new research avenue for the self-configuration and sustainable regulation of DMAs facing seasonal and foreseen water issues affecting the whole WDS but also unexpected, local undesirable events.

\end{document}